\documentclass[manuscript]{acmart}

\usepackage{comment}
\usepackage{color}
\usepackage{longtable}
\usepackage{xcolor}

\newcommand{\hl}[1]{#1}
\newcommand{\mytc}[1]{#1}

\AtBeginDocument{%
  \providecommand\BibTeX{{%
    \normalfont B\kern-0.5em{\scshape i\kern-0.25em b}\kern-0.8em\TeX}}}

\setcopyright{acmcopyright}
\copyrightyear{2018}
\acmYear{2018}
\acmDOI{XXXXXXX.XXXXXXX}
\begin{document}

\title{Gender Biases in Error Mitigation by Voice Assistants}

\author{Amama Mahmood}
\email{amama.mahmood@jhu.edu}

\affiliation{%
  \institution{Johns Hopkins University}
  \streetaddress{3400 N. Charles St}
  \city{Baltimore}
  \state{Maryland}
  \country{USA}
  \postcode{21218}
}

\author{Chien-Ming Huang}
\email{chienming.huang@jhu.edu}
\affiliation{%
  \institution{Johns Hopkins University}
  \streetaddress{3400 N. Charles St}
  \city{Baltimore}
  \state{Maryland}
  \country{USA}
  \postcode{21218}
}

\renewcommand{\shortauthors}{Mahmood and Huang}

\begin{abstract}

Commercial voice assistants are largely feminized and associated with stereotypically feminine traits such as warmth and submissiveness. As these assistants continue to be adopted for everyday uses, it is imperative to understand how the portrayed gender shapes the voice assistant's ability to mitigate errors, which are still common in voice interactions. We report a study (N=40) that examined the effects of voice gender (feminine, ambiguous, masculine), error mitigation strategies (apology, compensation) and participant's gender on people's interaction behavior and perceptions of the assistant. Our results show that AI assistants that apologized appeared warmer than those offered compensation. Moreover, male participants preferred apologetic feminine assistants over apologetic masculine ones. Furthermore, male participants interrupted AI assistants regardless of perceived gender more frequently than female participants when errors occurred. Our results suggest that the perceived gender of a voice assistant biases user behavior, especially for male users, and that an ambiguous voice has the potential to reduce biases associated with gender-specific traits.

\end{abstract}



\begin{teaserfigure}
  \includegraphics[width=\textwidth]{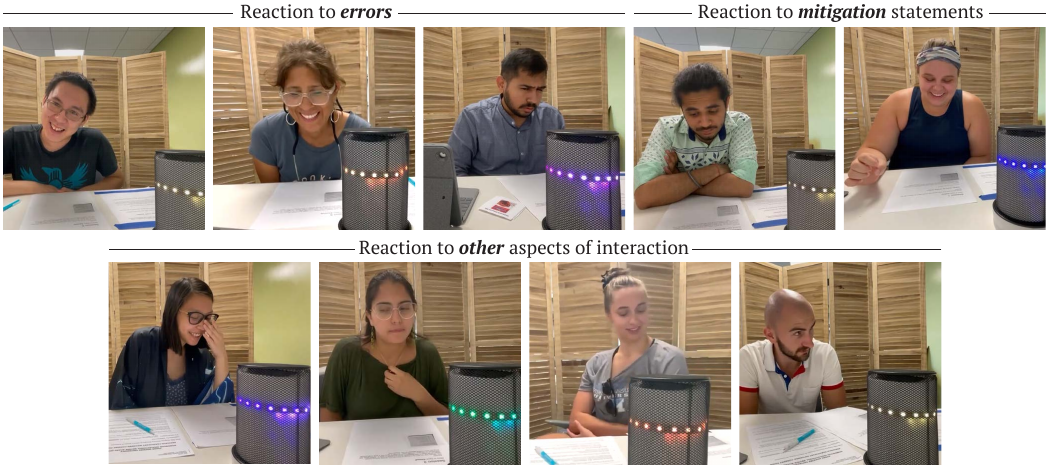}
  \caption{In this project, we investigate how the portrayed gender of a voice assistant, error mitigation strategy, and participant gender impact people's perceptions of and behavior towards the AI assistant in a collaborative task. 
  This figure illustrates that users react to errors made by voice assistants, to apology or compensation offered by voice assistants to mitigate negative effects of those errors, and to other aspects of interaction in general.}
  \Description[Voice assistants elicit non-verbal reactions from user in different conversational scenarios]{The figure consists of images of users reacting to three different conversational scenarios. There are total of 9 pictures. Each picture has a person sitting on a chair conversing with the smart speaker placed on a table in front of them. The The first three pictures show reactions of people to errors made by AI agent. The reactions are laughter by a male participant in the first picture, laughter by a female participant in the second picture, and a male participant frowning in the third picture. Next two pictures show reactions to mitigation statements that are; a frown by a male participant and laughter by a female participant. Next 4 pictures show reactions to other aspects of interactions which are; a female participant laughing, a female participant nodding, a female participant smiling, and a male participant raising eyebrows.}
  \label{fig:teaser}
\end{teaserfigure}

\maketitle

\section{Introduction}


Intelligent voice assistants are increasingly being used for a variety of assistive tasks in various personal and professional settings. 
For example, voice assistants, such as Alexa and Siri, are capable of supporting people while performing \textit{functional} tasks, such as retrieving information, setting alarms and reminders, sending texts, and making calls; and engaging in \textit{social} tasks, such as telling jokes, and playing music and games (e.g., ``Simon says''). Moreover, voice assistants are positioned to play an essential role in \mytc{future collaborative learning and work} \cite{baker2021intelligent, paraiso2006intelligent, saiz2020effectiveness, schmid2022does}. 

These modern voice assistants are powered by data-driven deep learning models to recognize speech commands and user intent to ensure smooth collaboration. 
While becoming powerful, these data-driven models are susceptible to errors \cite{pearl2016designing}, and when not mitigated and managed, these errors are likely to result in users distrusting and halting the collaboration. 
As a result, growing research has studied various strategies for mitigating AI errors.
For instance, \textit{apology} as an error mitigation strategy from an agent makes it likeable \cite{mahmood2022owning} and perceived as polite \cite{lee2010gracefully}, whereas \textit{compensation} increases users' satisfaction after error recovery \cite{lee2010gracefully}.

On top of experiencing imperfect interactions with voice assistants, users tend to personify \cite{habler2019effects} and characterize \cite{abercrombie2021alexa} these assistants as feminine \cite{un2019report}.
Such gender characterization is detrimental \hl{(for instance, it welcomes sexualised comments and insults }\cite{curry2019crowd,curry2020conversational}) because it fortifies women's amenable and unassertive role in the society \cite{loideain2020alexa, guzman2016making}.
Existing gender biases due to association of stereotypes in human-agent interactions \cite{ahn2022effect,lee2019voice} raise the following research questions that we aim to investigate: 
\hl{``\textit{how gender biases may potentially manifest when users interact with voice assistants, particularly in the context of erroneous interactions}''}
, and ``\textit{how the effectiveness of error mitigation strategies may be modulated by the portrayed gender of voice assistants}''.
The growing availability of (imperfect) voice assistants in digital services and the rooted gender biases in our society propel our inquiry into the relationships among the portrayed gender of a voice assistant, its mitigation strategy for errors, and the gender of the user when the user  is interacting  with a voice assistant. 

In this work, we experimentally manipulated \textit{the gender of voice assistant} (ambiguous, feminine, and masculine) and \textit{error mitigation strategy} (apology and compensation), and studied how these factors along with \textit{participant gender} (female (women) and male (men)) may influence people's perceptions of and behavior toward erroneous voice assistants.
We developed a mock smart speaker and conducted an in-person experiment where participants interacted with voice assistants to complete a series of shopping tasks (Figure \ref{fig:teaser}).
\hl{In this study, our emphasis is on the ``assistance'' scenarios, where commercial voice assistants excel in tasks such as online shopping, setting reminders, retrieving information, and creating lists. Therefore, we select the online shopping task as it exemplifies a typical assistance-type task performed by these voice assistants and has the potential for a longer, and multi-turn interaction between user and the voice assistant.}

Our experiment reveals several findings identifying nuanced dynamics among the portrayed gender of a voice assistant, mitigation strategy, and participant gender:
\begin{itemize}
    \item Participants reacted more socially, with a higher number of non-verbal reactions, to voice assistants that offer apology rather than compensation when mitigating errors.

    \item Male participants interrupted voice assistants, regardless of their portrayed gender and use of mitigation strategy, more often than female participants. 
    
    \item Male participants interacted more socially with ``apologetic'' voice assistants that were portrayed as feminine compared to their masculine counterparts. 
\end{itemize}
Overall, we also observed that male participants reacted more socially to the feminine assistants than masculine ones and that apology promotes the perception of warmth for voice assistants while compensation increases participants' satisfaction of service recovery, and perception of assistants' competence. 


\section{Background and Related Work}

\hl{Given that the primary focus of this paper is to explore gender biases, error mitigation strategies, and their intersection in user interactions with voice assistants, we first present related work on various error mitigation strategies for repairing human-agent relationships damaged by inevitable disruptions. We then provide an overview of previous studies that examine the stereotypes associated with gendered AI agents, with a specific emphasis on voice assistants.}


\subsection{Error Mitigation Strategies in Human-Agent Interaction}
It is crucial to repair relationships between \hl{agents\footnote{In our discussion of prior work, we have defined agents in a broad sense with respect to their embodiment. These agents encompass a wide range of forms, such as physical robots, AI aids, virtual avatars, voice assistants, and other similar entities.}} and humans after service breakdowns for continued use of technology. Mitigation strategies can help increase people's acceptance of faulty agents \cite{lee2010gracefully}. Moreover, these mitigation strategies can have varying effects on different dimensions of people's perceptions of the agents. For instance, a robot that apologizes for its mistake is considered more likeable and increases people's willingness to use it \cite{cameron2021effect}; however, it is not perceived equally capable to a robot that demonstrates the knowledge of why the error occurred and how the error can be fixed \cite{cameron2021effect}. Within apologies, a neutral apology, as opposed to humorous one, is found to be more sincere and effective in repairing human-agent relationship after errors made by smart speakers \cite{ge2019smart}. 
For voice-based interactions, combining acceptance of blame for the error and seriousness of tone while apologizing results in better user satisfaction after recovery from errors \cite{mahmood2022owning}; moreover, such agents are perceived to be more intelligent and likeable compared to the agents that lacked either seriousness or acceptance of blame \cite{mahmood2022owning}.

While apologies create an ``emotional shift toward forgiveness''  to promote reconciliation and regain trust during relationship repair \hl{in both human-human interactions} \cite{govier2002promise} \hl{and human-agent interactions} \cite{lee2010gracefully}, other mitigation strategies such as compensation may offer better user satisfaction while recovering from service failures. 
For instance, compensating with a free drink on a robot's failure to bring correct drink to the user resulted in higher service recovery satisfaction than acknowledging and apologizing for the error \cite{lee2010gracefully}. 
\hl{For deeper insights into the comparative aspects of apology and compensation strategies, we can look into the decades of research on mitigation strategies in repairing broken human-human 
relationships, which has a stronger foundation compared to research on human-agent interactions.}
\hl{For instance, human interactions in e-commerce show that the type of trust violation (competence- or integrity-based) also affects whether apologizing with internal or external attribution of blame, and compensating equal to or more than the loss would be more effective in regaining trust }\cite{cui2018use}.
Even in purely financial settings, apologies can facilitate preservation of relationships especially in the case of undercompensation that fails to redress the loss of the affected party; moreover, not surprisingly, overcompensation also repairs the relationship better than under or equal compensation \cite{haesevoets2013money}.  
\hl{Since in human-human interactions, apology is identified as a relational while compensation is termed as a financial (utilitarian) strategy}  \cite{haesevoets2013money}, \hl{we may find similar preferences towards apologies and compensation offered by intelligent agents.}
\hl{In the domain of human-robot interaction, we see that users' } orientation towards service can affect how the error mitigation strategies are perceived. People who want to maintain amicable relationship with service providers (stronger \textit{relational} orientation) prefer the apology strategy and have as much of negative impression for compensation strategy as for no strategy \cite{lee2010gracefully}. However, people who care more about quality of service than relationship with providers (stronger \textit{utilitarian} orientation) find the compensation strategy to be more appealing \cite{lee2010gracefully}. 
\hl{However, it is unclear how the preference towards apology and compensation might hold for voice assistants that do not have anthropomorphic physical embodiment.}
\hl{In this work, we seek to understand} how these two mitigation strategies compare in a voice-based interactions \hl{contextualized in an online shopping task} using a smart speaker.


\subsection{Gender Stereotypes in Human-Agent Interaction}
\hl{This section aims to provide an overview of the relevant literature on gender stereotypes and biases during human-agent interactions; in particular, we review the stereotypes and biases in 1) the broader literature on human-agent interactions, and 2) specific research focused on voice assistants. Additionally, we discuss the research regarding the utilization of different voice variations for assistants as a means of combating gender stereotypes and biases.}

\subsubsection{Gender Stereotypes in Human-Robot and Human-AI Interactions}
Design of ``masculine'' and ``feminine'' AI agents and robots may introduce gender based stereotypes. 
\hl{The terms ``masculine'' or ``feminine'' when referring to agents and robots encompass various methods that may contribute to their gendering. These methods may include manipulating their appearance, voice, and the use of specific names and pronouns in experiment instructions and surveys. Such gendering can be influenced by one or a combination of these factors.}
\hl{Thus, when we refer to an agent as ``masculine'' or ``feminine'', it means an agent that may be perceived to look, sound or act as a stereotypical male or female.}

\hl{Prior work in HRI supports that} masculine robots are considered better at stereotypical male tasks while feminine ones are perceived to be friendlier and warmer \cite{nomura2017robots}.  
\hl{Even with minimal gender markers, such as their voices or names, people tend to react in alignment with gender stereotypes seen in human interactions. For instance,}
\hl{robots with feminine names or voices} are expected to be emotionally intelligent \cite{chita2019gender}. 
\hl{Additionally, personality of the agent, perceived as feminine or masculine,  influences perception of trustworthiness, reliability and competence. Context and task at hand further shape how the gendered agents are perceived in these dimensions.}
\hl{For instance, a robot with a feminine personality reflected through agreeable and warm statements may be considered less trustworthy, reliable, and competent as opposed to one with masculine personality that is dominant, confident, and assertive, in a male-oriented task (e.g., dispatching taxis} \cite{kraus2018effects}).

\hl{The perception of individuals' competence is often associated with factors such as intelligence, efficiency, and skill, while warmth is more related to friendliness, sincerity, and helpfulness} \cite{fiske2007universal}. 
\hl{ Thus, considering masculine agents to be more competent while assigning higher warmth to feminine agents } \cite{ahn2022effect}\hl{, depicts association of gender stereotypes based on agent characteristics. 
Such biased perceptions also have implications for consumer preferences when interacting with AI agents in relation to different types of products. For example, }
in a chatbot-based interaction, masculine AI is preferred while recommending utilitarian products  \cite{ahn2022effect} (such as headphones for work from home)\hl{, which are practical and require competence from the seller to successfully convince buyers of their value} \cite{bennett2012universality}.
\hl{On the contrary, feminine AI fairs better in promoting hedonic products (related to happiness: drawing supplies)} \cite{ahn2022effect}\hl{, where warmth in the salesperson's personality influences purchasing decisions }\cite{bennett2012universality}.

\hl{In summary, various factors such as agent appearance, voice, personality, task, and context may significantly influence how the perceived gender of an agent may cause stereotyping. For this study, our focus is primarily  on examining the interaction between gendered voices of AI agents and error mitigation strategies in assistance-type task scenarios.}
We aim to understand how apology (relational/hedonic strategy) and compensation (financial/utilitarian strategy) interact with gendered voices of AI agents to influence user's perceptions. 

\subsubsection{Stereotypes Associated with Gendering of Voice Assistants}

Most voice assistants in our present digital world are gendered as female; for example, Apple's Siri, Amazon's Alexa, and Google voice assistant all have feminine voices set as default. 
These agents are intentionally feminized and marketed as ``fembots'' to increase their familiarity among users \cite{phan2019amazon,strengers2021smart,woods2018asking}.
Gendering of commercial voice assistants  poses societal harm because association of female gender by default to these personal voice assistants reinforces women's submissive role of taking and fulfilling orders in the society and welcomes harassment, abuse, and discriminatory behavior \cite{loideain2020alexa,guzman2016making,curry2020conversational,curry2019crowd}. 
In recent work, concerns about stereotypical assignment of female gender to agent voices have been raised, and power dynamic between the user and the woman-like voice assistants have been discussed \cite{hwang2019sounds}. 
A report by United Nations (UN) \cite{un2019report} states that despite what the creators claim, \textit{``nearly all of these assistants have been feminized---in name, in voice, in patterns of speech and in personality. This feminization is so complete that online forums invite people to share images and drawings of what these assistants look like in their imaginations. Nearly all of the depictions are of young, attractive women.''} 


\textbf{\hl{The influence of context and task on gender biases. }}
Prior research has yielded mixed results on the influence of task on gender biases in voice assistants.
On one hand, gender biases exist in the preferences for voice assistants and are dependent on contextual settings and task at hand. 
\hl{For example, a masculine voice is often favored for informative tasks, while a feminine voice is preferred for social interactions} 
\cite{lee2019voice}.
Similarly, feminine voice assistant gained more trust
in automation system targeted for home while masculine voice assistant gained more trust in office setting \cite{damen2019designing}. 
\hl{Moreover, in health-critical situations, users exhibit greater trust in feminine voice assistants compared to their masculine counterparts}
\cite{goodman2023s}. 
However, on the other hand, type of task may not always affect the perception of voice assistants. For instance, \hl{no gender biases were seen} while performing functional (information retrieval and mathematics calculations) and social (humours questions) tasks with Siri
\cite{lee2021social}. 
\hl{Such contradictory evidence on how task, especially assistive tasks such as controlling automated devices} \cite{damen2019designing} \hl{and retrieving information} \cite{lee2021social}, may affect the perception of voice assistants \hl{motivated our choice of an assistance-type task---online shopping---in our investigation of gender biases in error mitigation by voice assistants.}

\textbf{\hl{The influence of participant gender on gender biases. }}
\hl{In addition to context and task, user's own gender may shape their perceptions of voice assistants.}
\hl{For instance, prior work has demonstrated effect of participant's gender on their}
perception of trust \cite{lee2021social}.  
\hl{Participants tend to trust voices of the same gender more, even when uncertain about the accuracy of the advice or suggestion being given by voice assistants such as Siri} \cite{lee2021social}. 
\hl{Additionally, female participants have shown a higher level of skepticism compared to male participants, regardless of whether the voice they were interacting with belonged to a masculine or feminine assistant}
\cite{lee2021social}. 
\mytc{On the contrary, when interacting with animated pedagogical agents, female students showed a preference for a matched gender, whereas male students rated the female agent more favorably } \cite{ozogul2013investigating}. \mytc{Similarly, other studies indicate varied gender preferences} \cite{kramer2016closing, kim2007pedagogical}\mytc{, suggesting existence of potential gender stereotypes and biases.} 
Therefore, in this work, we also look at the role of gender of users, in addition to gender of assistants, in association of stereotypes with voice assistants. 


\subsubsection{Intervention for Mitigating Gender Stereotypes and Biases}
\hl{To address the issue of stereotypes and biases associated with gendered voices, researchers have explored various methods to mitigate their influence. 
One approach discussed to mitigate or eliminate the association of stereotypes with gendered voices is the creation of ambiguous voices through the manipulation of voice characteristics, such as pitch.}
Research has shown that there were no discernible differences in user trust formation between voice assistants with gendered voices, and those with claimed ambiguous voices 
\cite{tolmeijer2021female}; hence, ambiguous voices can potentially be used commercially in place of gendered assistants as a tool to reduce biases. 
\hl{However, it is worth mentioning that the gender ambiguous voice employed in the aforementioned study is unable to completely eradicate implicit biases, which are still evident similar to the biases associated with the feminine voice}\cite{tolmeijer2021female}, drawing our attention to lack of empirical evidence for effectiveness of ambiguous voices to culminate gender biases, implicit or explicit. 
\hl{Hence, in this work, in addition to masculine and feminine sounding assistants, we also study effects of ambiguous sounding assistants on people's perceptions of voice assistants in an online shopping task using a smart speaker.}

\subsubsection{Gaps in Methodological Approach  for Empirical Studies on Gender Biases}
\hl{Within the context of empirical studies on gender biases, a significant gap exists in the methodological approach employed.}
Most of prior studies on exploration of gender stereotypes and biases in voice assistants are based on online reviews \cite{phan2019amazon}, crowd sourced evaluations \cite{curry2019crowd}, case studies \cite{woods2018asking}, surveys \cite{obinali2019perception, curry2020conversational},  chat bots, and online interfaces \cite{damen2019designing, tolmeijer2021female, goodman2023s}.  
\hl{While there are a few exceptions, such as studies conducted with phone-based voice assistant Siri } \cite{lee2021social},
\hl{ or using off-the-shelf speakers controlled by experimenters} \cite{habler2019effects}
\hl{, these studies primarily depend on subjective evaluations through questionnaires to gain insights into stereotypes.}
\hl{In our work, we strive to address the methodological gap by introducing a simulated physical smart speaker platform through the implementation of the Wizard of Oz (WoZ) paradigm for the voice assistant. Moreover, we augment the subjective measures with an analysis of users' behavioral data to investigate implicit expression of biases associated with gender stereotypes.}

\subsection{Hypotheses}
\hl{Based on the prior work on mitigation strategies to repair human-agent relationship, and understanding various factors that impact gender stereotypes in human-agent interaction outlined above, we propose the following hypotheses:}
\begin{itemize}
    \item \textbf{Hypothesis 1.} Compensation, as a mitigation strategy, improves perceptions of service recovery satisfaction and competence more than apology while apology increases perceptions of warmth of the assistants. Hypothesis 1 is informed by previous research showing that 1) compensation is a financial strategy while apology is a relational strategy \cite{haesevoets2013money}, and 2) apologies make an agent or robot more likeable \cite{cameron2021effect} and appear more polite \cite{lee2010gracefully}, while compensation results in better service recovery satisfaction \cite{lee2010gracefully}. 
    \item \textbf{Hypothesis 2.} Feminine sounding assistants are perceived warmer than masculine sounding assistants whereas masculine assistants are considered more competent. Hypothesis 2 is informed by previous work showing that 
    masculine AI agent is rated more competent 
    while feminine AI agent is more likeable and warm \cite{ahn2022effect,nomura2017robots,chita2019gender}. 
    In addition to testing this hypothesis, we will explore if the gender ambiguous voice can culminate these biases. 
    \item \textbf{Hypothesis 3.} Assistants with feminine voice gain better user perceptions when they apologize for their mistake while assistants with masculine voice are perceived more positively when they compensate for their mistake. 
    Hypothesis 3 is formulated based on evidence from prior work in human-human interactions showing that  1) typically women are more apologetic than men \cite{turiman2013men}, 
    2) men are characterized to have authoritative and practical roles while women are considered better at submissive roles \cite{powell2002gender, koenig2011leader}; and 3) apology is a relational strategy while offering compensation is a financial (utilitarian) strategy \cite{haesevoets2013money} and may be considered as a feminine and a masculine trait, respectively. 
    Moreover, we explore where gender ambiguous voices would fall on this spectrum and if ambiguous assistants can reduce gender biases faced by AI agents while still being perceived as effective. 
    
\end{itemize}


Lastly, as discussed above, prior work on how participant gender may influence perception of masculine or feminine agents and robots has shown conflicting findings \cite{lee2021social, mitchell2011does, siegel2009persuasive}. Thus, in this study we also consider possible influence of participants' gender on their perceptions of gendered voice assistants while mitigating AI errors. \hl{Moreover, we conduct this study in the context of an assistance-type task and contextualize our results in that space.}

\section{Methods}\label{sec:methods}

\subsection{Study Design and Experimental Task}
To test our hypotheses, we conducted an in-person experiment considering three factors: \textit{error mitigation strategy}---(apology vs. compensation), \textit{gendered voice of AI assistants}---(ambiguous vs. feminine vs. masculine), and \textit{participant gender}---(female vs. male). Error mitigation strategy and gendered voice of AI assistants were within-subjects factors while participant gender was a between-subjects factor.

Our experiment consisted of six online shopping tasks; for each task, participants were asked to collaborate with one of the six AI voice assistants to complete the task.
We presented the six tasks as beta testing of a voice assistant developed for our prototype smart speaker. Each task involved working with a different voice assistant on a different shopping list. The six AI assistants were named Galaxy, Dot, Cloud, Mist, Phoenix, Gamma. We picked non-human and gender neutral names to reduce any biases in participant's preconception of the assistant's gender. The order of the assistant names and shopping lists was same for all the participants. For instance, the assistant in the first session is always named Galaxy and the participant ordered items for a party. Balanced Latin square is used to counterbalance the six experimental conditions. 

\begin{figure*}[t]
     \includegraphics[width=\textwidth]{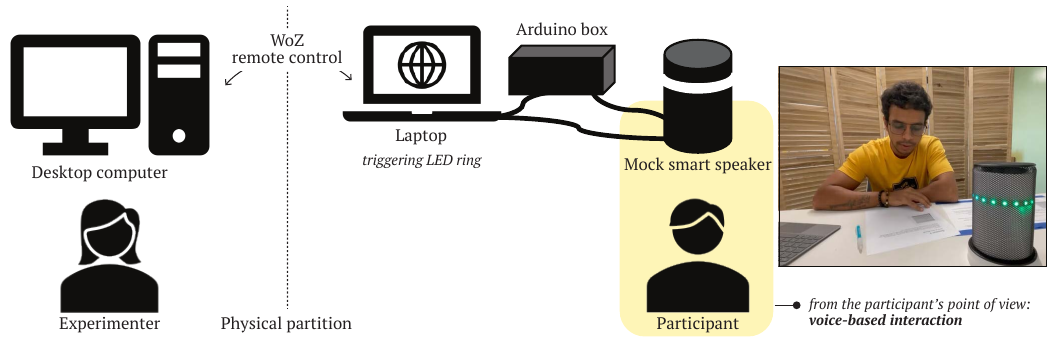}    
     \caption{Experimental setup. The participant interacts with the smart speaker using voice commands. The wired speaker is connected to the laptop. The LED lights ring on the speaker is activated by the Arduino once either of the two applications running on the laptop signals that assistant or participant speech is detected. The experimenter accesses the web interface on the laptop to control the speech of the mock smart speaker using TeamViewer on the Desktop (WoZ). The participant and the experimenter are in the same room separated by a physical partition.}
     \Description[Experimental set up explaining WoZ interface and interaction with user.]{The figure shows our experimental set up. To the left of the figure, the experimenter is shown with the desktop computer. To the right of a physical partition which separates the participant from the experimenter, the participant is shown with a wired speaker connected to Arduino and the laptop. To the very right, a picture of actual participant interacting with the smart speaker is shown where the person is sitting in front of the speaker which is placed on a table. The WoZ remote control is indicated between the desktop and the laptop. The experimenter simulates the assistant speech by controlling the laptop next to the wired speaker remotely through the desktop computer.}
    
    \label{fig:setup}
\end{figure*}

In each task, the user was given a list of five items to order through a speech-based interaction with the AI assistant (Fig. \ref{fig:setup}). The user would prompt the assistant to order a specific item and confirm the addition of item to the cart once the assistant presents the user with the item it found. The assistant would let the user know after adding the item to the cart.
The user could also ask the assistant to show more options.
The AI assistant would make a mistake on either the second, third, or fourth item on the list. 
In this study, we considered \textit{intent recognition} errors, which are described as ``recognized but incorrectly'' \cite{pearl2016designing}. The predetermined errors in this study were the use of homonyms---words that sound the same but may mean different things. For example, ``flour'' could be heard as flour or flower. As there are only a few possible items, mostly two, associated with the chosen homonyms, the assistant recovered from the error in the second attempt by suggesting correct item. 

The user could report a mistake by verbally indicating that the item suggested by the assistant is not the correct one when the assistant asked if it should add the item to the cart.
On indication of error by the user, the assistant prompted them to try again.
On the second try, the assistant fixed the mistake by indicating the correct item.
After fixing the error, the assistant performed the recovery strategy by initiating either apology or offering compensation based on the experimental condition.
After ordering the five items on the list, the user was directed to solve a riddle while they ``waited for the items to arrive''. The experimenter handed over the ordered items printed on a piece of paper with title and images to the user while they are solving the riddle. The user was prompted to see if the ordered items were correct.
The user could start a return process with the assistant in case any of the arrived items was incorrect. The assistant would initiate the return and notify that the item was returned and refunded. It was up to the user if they wanted to order the required item after returning the incorrect one or not. The assistant either apologized or offered compensation based on experimental condition once the return process was complete. 
Next, we describe the implementation details of our mock smart speaker used in the experiment.

\subsection{Mock Smart Speaker Setup}
We developed a mock smart speaker (Fig. \ref{fig:setup}) that can be controlled by an experimenter through Wizard of Oz (WoZ) paradigm. 
The speaker consisted of following components:
\begin{itemize}
    
    \item Arduino controlled LED lights ring for signaling  when the assistant was speaking or listening. LED lights were activated for signaling assistant's speech as soon as the experimenter played any audio on behalf of the assistant. For listening, the the LED light ring was activated on wake words which were either the name of the assistant, ``Hello'', ``Hey'', or ``Hi'' for the first time the assistant is addressed. Once the interaction started, if any speech from participant was detected right after the wake work or the assistant's turn of speaking (assistant waits for the participant's response after talking every time), the LED ring lighted up indicating that the speaker was listening to them. Each assistant had a different colored ring (blue, purple, green, pink, orange, and yellow) which was decided by the row of Latin square the participant was assigned to. We coded this behavior using two applications in $C\#$; 1) to detect audio output from the assistant, and 2) to detect and recognize the participant's speech. On detection of either participant or assistant's speech (as described above), the respective application will send a signal to Arduino to activate the lights ring. Both applications are running on the laptop (see Fig. \ref{fig:setup}). 
    
    \item Wired speakers for playing audio were connected to the laptop (see Fig. \ref{fig:setup}. The laptop powered and controlled the wired speaker, and the Arduino. The laptop was controlled remotely by the experimenter using TeamViewer\footnote{\url{https://www.teamviewer.com/en-us/}} on a desktop computer. The laptop screen was closed so that it is not visible to the participant. An iPad is also placed next to the wired speaker for the participants to complete puzzles and surveys throughout the study.
    
    \item Wizard of Oz interface is used to control the smart speaker's audio. The interface website, hosted on GitHub, dictated the flow of experiment and was running on the laptop connected to the wired speakers.  
    The experimenter operated the website remotely using TeamViewer on the desktop computer to control the assistant on the mock smart speaker using WoZ paradigm.
    The experimenter could click on various buttons to activate assistant responses to the participant. Apart from standard responses for each item in each shopping list such as ``\textit{Okay, I found Duracell Double A batteries. Should I add it cart}'', ``\textit{Let's try that again}'' for fixing the error, and ``\textit{The item has been added to your cart}'', the web interface also had generic responses to handle a participant's divergent requests. For instance, statements such as ``\textit{I am doing alright. How I can help you today?}'' to respond to the participant's questions about how the assistant is doing. 
    The simple responses such as ``\textit{Yes}'', ``\textit{No}'', ``\textit{Hi}'', and ``\textit{How can I help you today?}''  were also included. For requests that could not be fulfilled by the speaker such as the participant asking to read out all the items in their cart, the assistant responded with ``\textit{Unfortunately, that is currently beyond my capabilities.}'' and presented a possible solution to continue the conversation, ``\textit{but you can ask me if a specific item is in the cart or not?}''. To handle such specific queries, we also added ``\textit{Yes, that item is already in the cart}'' and ``\textit{No, that item is not in the cart yet, would you like me to add it?}''. If the participant requests to remove any particular item from the cart, the assistant obliged and responded, ``\textit{The item has been removed from the cart.}'' If the participant forgot to ask the voice assistant to check out at the end of the task, the assistant prompted the participant by asking ``\textit{Would you like to check out?}'' On the participant's request or confirmation to check out, the assistant said ``\textit{Okay, checking out. Your items should arrive shortly}''. To handle return process, there were a few statements at the experimenter's disposal such as ``\textit{Okay, I can start a return process for the item}'', ``\textit{Which item would you like to return?}'', and ``\textit{Your item has been returned and a refund has been issued}'' in addition to the error mitigation statement. 
    
\end{itemize}

Next, we discuss design of our manipulations for error mitigation strategies and gendered voices.

\subsection{Manipulation of Error Mitigation Strategies}
In this study, we focused on two error mitigation strategies, namely apology and compensation; the strategy of apology has been employed in prior research with a similar online shopping context \cite{mahmood2022owning}, whereas the strategy of compensation is relevant given the nature of the shopping task.  
\begin{itemize}
    \item \textbf{\textit{Apology.}} The design of our apology strategy followed a similar design of error responses in a prior study on online shopping with a smart speaker in a simulated online experiment \cite{mahmood2022owning}; in particular, the study showed that an apology offered in a serious tone with the acceptance of blame for the AI error is preferred over the apologies that lacked seriousness, acceptance of blame, or both.
    In our study, the agent apologized for the AI error by saying ``I am sorry for the inconvenience. I confused the items because there are multiple items for this keyword. From time to time, I have difficulty distinguishing among such items.''
    
    \item \textbf{\textit{Compensation.}} For the compensation strategy, the agent offered a 10 dollar credit to the user by saying ``\textit{I will give you a 10 dollar credit that can be applied to your future purchases with us for the inconvenience caused.}''
\end{itemize}

\subsection{Manipulation of Gendered Voices}
We manipulated the gender of AI voice to sound feminine, masculine and ambiguous. We chose the feminine and masculine voices from two free text-to-speech websites ``Sound of Text''\footnote{\url{https://soundoftext.com/}} and ``wideo''\footnote{\url{https://wideo.co/text-to-speech/}} (US Jack Bailey), respectively. The voices had American accent, and default pitch and speed. Inspired from prior work \cite{tolmeijer2021female}, we generated gender ambiguous voices by shifting the pitch of the above feminine and masculine voices by 2 or 3 semitones where shifting by one semitone is equivalent to shifting by 5.946\%. We used an online pitch shifting tool\footnote{\url{https://onlinetonegenerator.com/pitch-shifter.html}}. The masculine voice was increased by 2 and 3 semitones while the feminine voice was lowered by 2 and 3 semitones to obtain 4 candidates for gender ambiguous voices. To select a gender ambiguous voice, we conducted two pilot studies:

\subsubsection{Pilot Study 1: Picking 2 gender ambiguous voices from 4 candidates}  
To pick one ambiguous voice between the shifted masculine voices and one between the shifted feminine voices, we conducted an online poll with nine participants who listened to the 4 voices and were asked to choose two ambiguous voices, one from each gender category.
Moreover, they were asked to rank the voices from most ambiguous to least ambiguous. Six of the 9 participants picked feminine-3 semitones to be more ambiguous among shifted feminine voices which is consistent with the results from prior study \cite{tolmeijer2021female}. Whereas, seven of the 9 participants picked the masculine+2 semitones as more ambiguous from the shifted masculine voices. Overall, masculine+2 semitones was picked by eight participants to be in top 2 when they were asked to rank the voices for most ambiguous to least ambiguous voices. While feminine-3 semitones was picked by four participants to be in top 2. Based on the results from this poll, we kept masculine+2 semitones (masculine-ambiguous) and feminine-3 semitones (feminine-ambiguous) as two candidates for gender ambiguous voice for the next pilot study.

\subsubsection{Pilot Study 2: Picking 1 gender ambiguous voice from 2 candidates and validating the manipulation of all gendered AI voices.} 
To verify that the gendered voices, especially ambiguous voices, are perceived as intended regardless of the nature of a task, 
we ran a within-subjects study with two factors: \textit{voice gender}---(feminine, masculine feminine-ambiguous, masculine-ambiguous) and \textit{nature of task}---(stereotypically masculine, neutral, stereotypically feminine). 
For the manipulation of task nature, we picked following three statements narrated by the voice agents:
\begin{itemize}
    \item \textbf{\textit{Stereotypically masculine task}}  (fixing a car). ``\textit{If the car’s charging port is not working, check to see if the fuse for the power outlet has blown}''.
    \item \textbf{\textit{Neutral task}} (defining a term). ``\textit{A recession is a period of temporary economic decline during which trade and industrial activity are reduced}''.
    \item \textbf{\textit{Stereotypically feminine task}} (nursing). ``\textit{The doctor will be with you shortly. Meanwhile, please fill out the consent forms}''.
\end{itemize}
\hl{We recruited 16 participants (gender: 9 female, 7 male; age: $M = 23.33, SD = 2.61$) through convenience sampling from the local community.}
The participants were presented with 12 conditions in a randomized order. In each condition, the participants listened to the voice of one of the AI agents saying one of the above statements and then rated the agent in terms of \textit{perceived gender} (1 being feminine, 3 being ambiguous and 5 being masculine) and \textit{humanlikeness} (1 being robotic and 5 being human) on 5-point scales.
Furthermore, participants were asked to give a name to the agent. Two coders independently assigned gender (masculine, feminine, or ambiguous) to the names provided by the participants. In case of disagreement between two coders, they discussed and reached a consensus on the perceived gender based on the name given to the agent. 

\hl{The data collected from the second pilot study was analyzed to select the voices for the voice assistants.} To validate the manipulation of perceived gender of the AI voice, we compared the rating of perceived gender to the baseline, the value 3 (ambiguous) on a 5-point scale. 
One-way analysis of variance (ANOVA) revealed that the AI voice had a significant main effect on perceived gender, $F(4,235) = 146.04, p<.001$. Pairwise comparisons using Dunnett's test with control (baseline: perceived gender = 3, ambiguous) revealed that all the voices except masculine-ambiguous ($M=3.25,SD=0.84$) is significantly different from the baseline. The masculine voice ($M=4.52, SD =0.74$) is significantly higher on perceived gender (5-point scale, 1 being feminine, 3 being ambiguous, and 5 being masculine) than baseline ($M=3, SD = 0$), $p<.001$ while the feminine voice ($M=1.14, SD=.36$), $p<.001$, and feminine-ambiguous ($M=2.65,1.02$), $p=.046$, voices are significantly lower on perceived gender than baseline. These results verified that our manipulation for masculine and feminine voices was adequate. 
A chi-square test of independence was performed to examine the effect of gender of AI voice and the perceived gender of the names given by participants. The relation between these variables was significant ($\chi^2(6, N=192) = 169.21, p < .001$). Our masculine-ambiguous AI voice was given gender ambiguous names ($66.7\%$) more often than the feminine-ambiguous AI voice ($62.5\%$).
Based on these results, we concluded that masculine-ambiguous voice was perceived to be more gender ambiguous than the 
feminine-ambiguous voice. As a result, we used the masculine-ambiguous voice as a gender ambiguous voice in our main study. 

Furthermore, on running two-way repeated measures ANOVA with voice gender and nature of task as factors, there is no significant main effect of nature of task on perceived gender. 
A two-way repeated measures ANOVA revealed no significant main effect of gender and nature of task on humanlikeness. However, there is a significant interaction effect between voice gender and nature of task, $F(6,90) = 2.46, p = .030, \eta^2 = .141$; Tukey's HSD post hoc test shows no significant pairwise comparisons.

\subsection{Measures}
For the main study, we used a range of metrics to measure user perceptions of agent characteristics, subjective evaluations of the voice assistants, and user behavior towards the assistants. 

\subsubsection{Perceptions of Agent Characteristics}
\begin{itemize}
    \item \textbf{Perceived gender}. To check if our manipulation of the gender of AI voices was perceived as intended, we asked the participants to rate the voice gender on a 5-point scale (1 being feminine, 3 being ambiguous, and 5 being masculine).
    \item \textbf{Humanlikeness}. We asked the participants to rate humanlikness of the agent on a 5-point scale (1 being robotic and 5 being human).

\end{itemize}

\subsubsection{Subjective Measures of User Experience and Perceptions of AI}

\begin{itemize}
    \item \textbf{Service recovery satisfaction} (Three items; Cronbach's $\alpha = .91$). 
    We used three questions (``\textit{The AI assistant was responsive to any errors.}'', ``\textit{I am happy with how the error was handled}'' and ``\textit{In my opinion, the AI assistant provided a satisfactory response to the error}'') as informed by prior work on apology \cite{roschk2013nature} in the domain of consumer services \cite{tax1998customer,maxham2003firms} to measure service recovery satisfaction. 

    \item \textbf{Warmth} (Four items; Cronbach's $\alpha = 87$). 
    We asked the participants to rate their impression of the agent on these dimensions: 1) Cold -- Warm, 2) Unfriendly -- Friendly, 3)  Insincere -- Sincere, and 4) Boastful -- Modest.

    \item \textbf{Competence} (Three items; Cronbach's $\alpha = .86$). 
    We asked the participants to rate their impression of the agent on these dimensions: 1) Unskillful -- Skillful, 2) Impractical -- Practical, and 3) Foolish -- Intelligent as informed by prior work on gender stereotypes on AI recommendations \cite{ahn2022effect} and social cognition dimensions \cite{fiske2007universal} similar to `warmth'.

\end{itemize}

\subsubsection{Behavioral Measures}
To understand participants' behaviors toward voice assistants during their interactions, we asked two human coders to annotate the interaction videos and extracted the following behavioral measures based on the annotation. The two coders annotated same 10\% of the participant videos for inter-coder reliability; the rest of the videos were 50:50 split between the two coders. For dichotomous variables, percentage agreement was reported for assessing inter-coder reliability. For continuous variables, Intraclass Correlation Coefficient (ICC) estimates and their 95\% confident intervals were calculated and reported using SPSS 
assuming single measurement, consistency, two-way mixed-effects model. ICC coefficients between 0.75 and 0.90, and greater than 0.90 are indicative of good and excellent reliability, respectively \cite{koo2016guideline}.
\begin{itemize}
    \item \textbf{Participant interruption during error (Dichotomous).}  This measure encodes whether the participant interrupted the assistant after realizing that an error had occurred while the assistant was still talking. 
    Essentially, this measure indicates if the participant started talking before the assistant stopped. For instance, during the statement with error, ``\textit{Okay, I found red archery bow. Should I add it to the cart?}'', the interruption would be recorded if the participant starts talking before the assistant finishes saying ``\textit{Should I add it to the cart?}''. There was a 100\% agreement between the two coders.
    \item \textbf{Non-verbal reaction to error (Dichotomous).} This measure encodes whether the participant reacted to the error non-verbally (e.g., changes in face and/or upper body; see Fig. \ref{fig:teaser} for example reactions). There was a 100\% agreement between the two coders.
    \item \textbf{Verbal response to mitigation statement (Dichotomous).} We look at whether the participant responds verbally to the error mitigation statement (compensation or apology). For instance, participants saying ``\textit{Thank you}'' and ``\textit{It's okay}'' would would count as verbal responses to error mitigation strategy. There was a 100\% agreement between the two coders.
    \item \textbf{Non-verbal response to mitigation statement (Dichotomous).} We look at whether the participant reacts non-verbally (changes in face and/or upper body; see Fig. \ref{fig:teaser} for example responses) to the error mitigation statement (compensation or apology). There was a 100\% agreement between the two coders.
    \item \textbf{Number of other interruptions (interruptions that were not concerned with the error).} Interruptions are defined as when the participant and the assistant interrupts or talks over each other. This measure counts the number of interruptions, apart form those to error as mentioned above, during the interaction, trying to capture the overall interaction fluency. 
    A good degree of reliability was found between 2 raters, $ICC(3,1) = .835 [.644, .928]$. 
    \item \textbf{Number of reactions (verbal and non-verbal)}. Verbal reactions are recorded when the participant says phrases other than direct response to the speaker such as ``\textit{Thank you}'', ``\textit{You're good}'' etc. Non-verbal reactions, such as smile, laugh, and raising eye brows, are recorded when the participant's face and/or upper body (visible in video recordings) changes as reaction to the assistant's speech (see Fig. \ref{fig:teaser} for example non-verbal reactions). Involuntary movements were not recorded as non-verbal reactions.
    We do not include verbal and non-verbal responses to error in this measure because we are interested in understanding, in general, when nothing unexpected occurs how social the interaction of the participant with the smart speaker is. Hence, we focus only on the conversational aspect of the interaction, which includes responses to mitigation statements. 
    A good degree of reliability was found between 2 raters for number of verbal reactions, $ICC(3,1) = .837 [.648, .929]$. Similarly, a good degree of reliability was also found between 2 raters for number of non-verbal reactions, $ICC(3,1) = .896 [.766, .955]$.

\end{itemize}

\subsection{Procedure}
This study consisted of five phases: 

\begin{enumerate}

\item \textit{Introduction and consent.} At the start of the study, participants were provided with a brief description of the study. The description stated that participants would be ordering items from shopping lists while collaborating with six AI assistants to test our smart speaker prototype. The participation was voluntary; participants agreed to continue the study by signing the consent form approved by our Institutional Review Board (IRB). 

\item \textit{Pre-study questionnaire.} The participants filled a pre-study questionnaire inspired from prior work \cite{bradley2016toward} to evaluate users' utilitarian or relational orientation regarding shopping experience.

\item \textit{Experimental task: simulated shopping.} 
Participants were assigned one of the rows in a Latin square of order 6, dictating the order of experimental tasks. They interacted with the AI assistant on the mock smart speaker.
Participants were prompted to do a puzzle while waiting for their order to arrive. After the experimenter handed them the items (printed on a piece of paper) that they ordered, participants verified if the items were correct, and started a return with the speaker if needed.

\item \textit{Perception survey.} After interacting with the AI assistant, participants completed a questionnaire about their perceptions of the AI assistant. They continued onto the next condition and repeated phases 3 and 4.

\item \textit{Post-study questionnaire.} After completing all the conditions, participants filled out a post-study demographics questionnaire.

\end{enumerate}

The study was approved by our IRB.
The study took approximately 45 minutes to complete. The participants were compensated with a \$15 USD gift card for their participation in the study. 

\subsection{Participants}
We recruited 43 participants using university mailing lists and flyers around the campus. Two of the participants were excluded from data analysis because they failed to identify the targeted error correctly in 5 out of 6 trials. A total of 41 participants (21 females (women), 19 males (men), 1 non-binary) were considered for data analysis. The participants were aged between 18 to 58 ($M=26.20, SD=7.92$) and had a variety of education backgrounds, including computer science, engineering and technology, healthcare, life sciences, media sciences, music, and education. Twenty seven participants indicated their ethnicity to be Asian, 2 African American, 6 Caucasian, 5 Spanish origin of any race, and 1 Caucasian and Spanish origin of any race. 

\begin{table}[tb]
\centering
\caption{Results of perceived gender of agent voice(1: feminine, 3: ambiguous, and 5: masculine) and humanlikeness. Mixed-model repeated measures ANOVA for voice gender, mitigation strategy, and participant gender.  $^{\ast}p \leq 0.05$, $^{\ast\ast}p \leq 0.01$, and  $^{\ast\ast\ast}p\leq 0.001$}
\label{tab:characteristics-perceptions-results}
\begin{tabular}{llcl}

\textbf{Perceived Voice Gender} (1: feminine 3: ambiguous 5: masculine) \\
\hline
Variables & $F$ & \textbf{$p$} & $\eta_{p}^{2}$ \\
\hline
\hline
Voice Gender    & $F(2,76) = 323.55$ & $<.001^{\ast\ast\ast}$  & $.895$ \\
\hspace{3mm} Feminine ($M= 1.14, SD=0.44$) < Masculine  ($M= 4.89, SD=0.34$) & & $<.001^{\ast\ast\ast}$\\
\hspace{3mm} Feminine ($M= 1.14, SD=0.44$) < Ambiguous  ($M= 3.93, SD=1.18$) & & $<.001^{\ast\ast\ast}$\\
\hspace{3mm} Ambiguous  ($M= 3.93, SD=1.18$) < Masculine ($M= 4.89, SD=0.34$) & & $<.001^{\ast\ast\ast}$\\
\hline
Mitigation Strategy    & $F(1,38) = 0.15$ & $.700$ & $.004$  \\
\hline
Participant Gender     & $F(1,38) = 5.96$ & $.019^{\ast}$ & $.136$ \\
\hspace{3mm} Male ($M= 3.17, SD=1.82$) < Female  ($M= 3.45, SD=1.70$) & & \\
\hline
Participant Gender $\times$ Voice Gender  & $F(2,76) = 2.86$ & $.063$  & $.070$ \\
\hline
Participant Gender $\times$ Mitigation Strategy  & $F(1,38) = 3.77$ & $.072$ & $.083$ \\
\hline
Voice Gender $\times$ Mitigation Strategy  & $F(2,76) = 1.69$ & $.192$ &  $.043$ \\
\hline
Participant Gender $\times$ Voice Gender \& Mitigation Strategy  & $F(2,76) = 0.52$ & $.598$ & $.013$\\
\hline
\\
\textbf{Humanlikeness} \\
\hline
Variables & $F$ & \textbf{$p$} & $\eta_{p}^{2}$ \\
\hline
\hline
Voice Gender    & $F(2, 76) = 21.18$ & $<.001^{\ast\ast\ast}$  & $.358$ \\
\hspace{3mm} Masculine  ($M= 2.96, SD=1.06$) < Feminine ($M= 3.36, SD=1.02$)    & & $.003^{\ast\ast}$\\
\hspace{3mm} Ambiguous  ($M= 2.59, SD=1.13$) < Feminine ($M= 3.36, SD=1.02$)    & & $<.001^{\ast\ast\ast}$\\
\hspace{3mm} Ambiguous  ($M= 3.93, SD=1.18$) < Masculine ($M= 2.96, SD=1.06$)   & & $.007^{\ast\ast}$\\
\hline
Mitigation Strategy    & $F(1,38) = 0.21$ & $.646$ & $.006$  \\
\hline
Participant Gender     & $F(1,38) = 0.90$ & $.348$ & $.023$ \\
\hline
Participant Gender $\times$ Voice Gender  & $F(2,76) = 1.26$ & $.291$  & $.032$ \\
\hline
Participant Gender $\times$ Mitigation Strategy  & $F(1,38) = 3.56$ & $.067$ & $.086$ \\
\hline
Voice Gender $\times$ Mitigation Strategy  & $F(2,76) = 0.48$ & $.620$ &  $.013$ \\
\hline
Participant Gender $\times$ Voice Gender \& Mitigation Strategy  & $F(2,76) = 0.54$ & $.583$ & $.014$\\
\hline
\end{tabular}
\end{table}

\section{Results}\label{sec:results}
Our data analysis included a total of 246 trials from 41 participants (six trials for six conditions per participant).
In 9 of the 246 trials, the intended errors were not correctly identified. To handle these missing values for the questionnaire measures, first, we analyzed whether these values are missing completely at random (MCAR) or not. Little's MCAR test \cite{little1988test} was not significant suggesting that it is safe to assume that data is MCAR, $\chi^2 (197,N=43) = 158.28$, $p = .98$.
Multiple Imputation (MI) \cite{rubin2004multiple} is one of the optimal techniques to handle the missing data and gives unbiased results under MCAR \cite{van2020rebutting}. Thus, before proceeding with our analysis, we replaced the missing data using MI in SPSS by computing five imputations and pooling the results, taking into account variation across these imputations. 

For the analyses reported in this paper, we included participant gender as a between-subjects factor to explore the nuanced relationships among voice gender, mitigation strategy, and participant gender. Due to the significant imbalance in participant gender, with only one participant who self reported having a non-binary gender, we excluded this participant in the following analyses. Thus, a total of 40 participants (21 females (women), 19 males (men)) aged between 18 to 58 ($M=26.23, SD=8.08$) were included for analysis.
For the results reported below, we used mixed-model repeated measures analysis of variance (ANOVA) unless stated otherwise. The gender of AI voice, error mitigation strategy, and participant's gender were set as the fixed effect and participants as a random effect. Voice gender and error mitigation strategy are within-subjects factor while participant gender is between-subjects factor. 
For all the statistical tests reported below, $p<.05$ is considered as a significant effect. 
\hl{We only describe significant results below. Complete results are presented in Tables } \ref{tab:characteristics-perceptions-results}, \ref{tab:subjective-results}, \ref{tab:behavioral-1}, and \ref{tab:behavioral-2}. 
We follow Cohen's guidelines on effect size and considered  $\eta_p^2=0.01$ a small effect, $\eta_p^2=0.06$ a medium effect, and $\eta_p^2=0.14$ a large effect \cite{cohen1988statistical}.

\begin{figure*}[tb]
    \includegraphics[width=\textwidth]{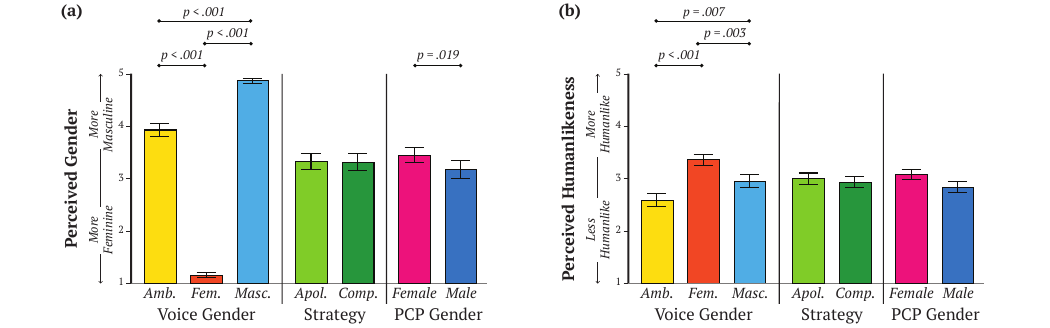}
    \caption{Results of perceived gender and humalikeness of the assistant. Mixed-model repeated measures ANOVAs were conducted to discover effects of voice gender---ambiguous (Amb.), feminine (Fem.), masculine (Masc.)---, mitigation strategy---apology (Apol.), compensation (Comp.)---, and participant (PCP) gender---female, male--- on participants' perceptions of agent characteristics. All pairwise comparisons were conducted using Fisher's LSD method with Bonferroni correction. Error bars represent standard error (SE) and only the significant comparisons ($p<.05$) are highlighted.}
    \label{fig:gender}
    \Description[Results of perceived gender and humalikeness of the assistant.]{The figure shows results of perceived gender and humalikeness of the assistant. The figure further consists of 2 sub graphs. All are bar plots and y-axis is ranged from 1 to 5. 
    The first of the 2 sub graphs from left to right further consists of 3 bar plots and shows results on perceived gender which is the y-axis for all three bar plots. One means most feminine and 5 means most masculine while 3 means ambiguous. 
    First of the 3 bar plots has voice gender on x-axis and shows significant main effect of voice gender on perceived gender of assistants. The p-value of all the significant comparisons discussed is less than .001 unless stated otherwise. Feminine assistant with a mean of 1.1 is rated more feminine than masculine assistant with a mean of 4.9. While, ambiguous with mean of about 3.9 is also rated more masculine than feminine and more feminine than masculine assistant. 
    Second of the 3 bar plots has mitigation strategy on x-axis and shows no significant  main effect of mitigation strategy on assistant's perceived gender. Assistants employing compensation and apology strategy have mean of about 3.5. Third of the 3 bar plots has participant gender on x-axis and shows significant effect of participant gender on assistants' perceived gender with a p-value of.019. Male participants with a mean of 3.2 perceived the assistants' gender to be lower on the scale which is more feminine than female participants with a mean of 3.4. 
    The figure shows results of perceived gender and humalikeness of the assistant. The figure further consists of 2 sub graphs. All are bar plots and y-axis is ranged from 1 to 5. 
    The second of the 2 sub graphs from left to right further consists of 3 bar plots and shows results on perceived humanlikeness which is the y-axis for all three bar plots. One means least humanlike and 5 means most humanlike. 
    First of the 3 bar plots has voice gender on x-axis and shows significant main effect of voice gender on perceived gender of assistants. Feminine assistant with a mean of 3.4 is rated more humanlike than masculine assistant with a mean of 3.0 and a p-value of .003. While, ambiguous with mean of about 2.6 is rated less humanlike as compared to masculine with a p-value of .007 and feminine assistants with a p-value of less than .001. 
    Second of the 3 bar plots has mitigation strategy on x-axis and shows no significant  main effect of mitigation strategy on assistant's perceived humanlikeness. Assistants employing compensation and apology strategy have mean of about 2.9 and 3.0 respectively. 
    Third of the 3 bar plots has participant gender on x-axis and shows no significant  main effect of  participant gender on assistants' perceived humalikeness. Male participants have a mean of 3 and female participants have a mean of  3.2.
    }
\end{figure*}

\subsection{Perceptions of Agent Characteristics}
Fig. \ref{fig:gender} and \hl{Table } \ref{tab:characteristics-perceptions-results} summarize our results for perceptions of agent characteristics. 

\subsubsection{Perceived Voice Gender} 
\hl{We saw a significant main effect of the voice gender on participants' perception of agent's gender confirming that the  manipulation of gendered voices for assistants in the study was successful} (Fig. \ref{fig:gender} a and Table \ref{tab:characteristics-perceptions-results}).
\hl{The agent with feminine voice  was rated more feminine than the agents with masculine and ambiguous voice. Masculine voice was rated more masculine than ambiguous voice. 
Moreover, we observed a significant main effect of participant gender on their perceptions of agent's gender. Male participants perceived assistants' gender to be lower on the scale (more feminine) than female participants}.

\subsubsection{Humanlikeness}
\hl{We found a significant main effect of voice gender on participants' perception of humanlikeness of the agent} (Fig. \ref{fig:gender} b and Table \ref{tab:characteristics-perceptions-results}). 
\hl{The agent with feminine voice was perceived more humanlike than the agent with masculine voice. Ambiguous voice was perceived more robotic than both feminine and masculine voice.}

\begin{table}[htbp]
\centering
\caption{Results of subjective measures: service recovery satisfaction, warmth, and competence. Mixed-model repeated measures ANOVA for voice gender, mitigation strategy, and participant gender.  $^{\ast}p \leq 0.05$, $^{\ast\ast}p \leq 0.01$, and  $^{\ast\ast\ast}p\leq 0.001$}
\label{tab:subjective-results}
\begin{tabular}{llll}

\textbf{Service Recovery Satisfaction}\\
\hline
Variables & $F$ & \textbf{$p$} & $\eta_{p}^{2}$ \\
\hline
\hline
Voice Gender    & $F(2,76) = 0.99$ & $.377$  & $.025$ \\
\hline
Mitigation Strategy    & $F(1,38) = 4.81$ & $.034^{\ast}$ & $.112$  \\
\hspace{3mm} Apology ($M= 4.22, SD=0.84$) < Compensation  ($M= 4.47, SD=0.76$) & &\\
\hline
Participant Gender     & $F(1,38) = 1.93$ & $.173$ & $.048$ \\
\hline
Participant Gender $\times$ Voice Gender  & $F(2,76) = .310$ & $.720$  & $.039$ \\
\hline
Participant Gender $\times$ Mitigation Strategy  & $F(1,38) = 3.77$ & $.060$ & $.090$ \\
\hline
Voice Gender $\times$ Mitigation Strategy  & $F(2,76) = 0.19$ & $.831$ &  $.005$ \\
\hline
Participant Gender $\times$ Voice Gender \& Mitigation Strategy  & $F(2,76) = 0.29$ & $.712$ & $.008$\\
\hline
\\
\textbf{Warmth}\\
\hline
Variables & $F$ & \textbf{$p$} & $\eta_{p}^{2}$ \\
\hline
\hline
Voice Gender    & $F(2,76) = 3.93$ & $.024^{\ast}$  & $.094$ \\
\hspace{3mm} Ambiguous ($M= 3.79, SD=0.82$) < Feminine  ($M= 4.03, SD=0.70$) & &$.027^{\ast}$&\\
\hline
Mitigation Strategy    & $F(1,38) = 6.39$ & $.016^{\ast}$ & $.144$  \\
\hspace{3mm} Compensation ($M= 3.80, SD=0.77$) < Apology  ($M= 3.98, SD=0.74$) & &\\
\hline
Participant Gender     & $F(1,38) = 0.96$ & $.334$ & $.025$ \\
\hline
Participant Gender $\times$ Voice Gender  & $F(2,76) = 1.52$ & $.225$  & $.039$ \\
\hline
Participant Gender $\times$ Mitigation Strategy  &$F(1,38) = 0.05$ & $.826$ & $.001$ \\
\hline
Voice Gender $\times$ Mitigation Strategy  & $F(2,76) = 0.42$ & $.656$ &  $.011$ \\
\hline
Participant Gender $\times$ Voice Gender \& Mitigation Strategy  & $F(2,76) = 1.953$ & $.149$ & $.023$\\
\hline
\\
\textbf{Competence}\\
\hline
Variables & $F$ & \textbf{$p$} & $\eta_{p}^{2}$ \\
\hline
\hline
Voice Gender    & $F(2,76) = 3.56$ & $.033^{\ast}$  & $.086$ \\
\hspace{3mm} Masculine ($M= 3.83, SD=0.85$) < Feminine  (($M= 4.00, SD=0.84$)) & & $.021^{\ast}$&\\
\hline
Mitigation Strategy    & $F(1,38) = 6.39$ & $.016^{\ast}$ & $.144$  \\
\hspace{3mm} Compensation ($M= 3.80, SD=0.77$) < Apology  ($M= 3.98, SD=0.74$) & &\\
\hline
Participant Gender     & $F(1,38) = 1.93$ & $.173$ & $.048$ \\
\hline
Participant Gender $\times$ Voice Gender  & $F(2,76) = 1.20$ & $.308$  & $.030$ \\
\hline
Participant Gender $\times$ Mitigation Strategy  &$F(1,38) = 0.00$ & $.997$ & $.000$ \\
\hline
Voice Gender $\times$ Mitigation Strategy  & $F(2,76) = 0.18$ & $.794$ &  $.005$ \\
\hline
Participant Gender $\times$ Voice Gender \& Mitigation Strategy  & $F(2,76) = 0.29$ & $.712$ & $.008$\\
\hline
\end{tabular}

\end{table}

\subsection{Subjective Measures}
Fig. \ref{fig:service} and \hl{Table } \ref{tab:subjective-results} summarize our results for subjective measures.  

\subsubsection{Service Recovery Satisfaction}
\begin{figure*}[t]
    \includegraphics[width=\textwidth]{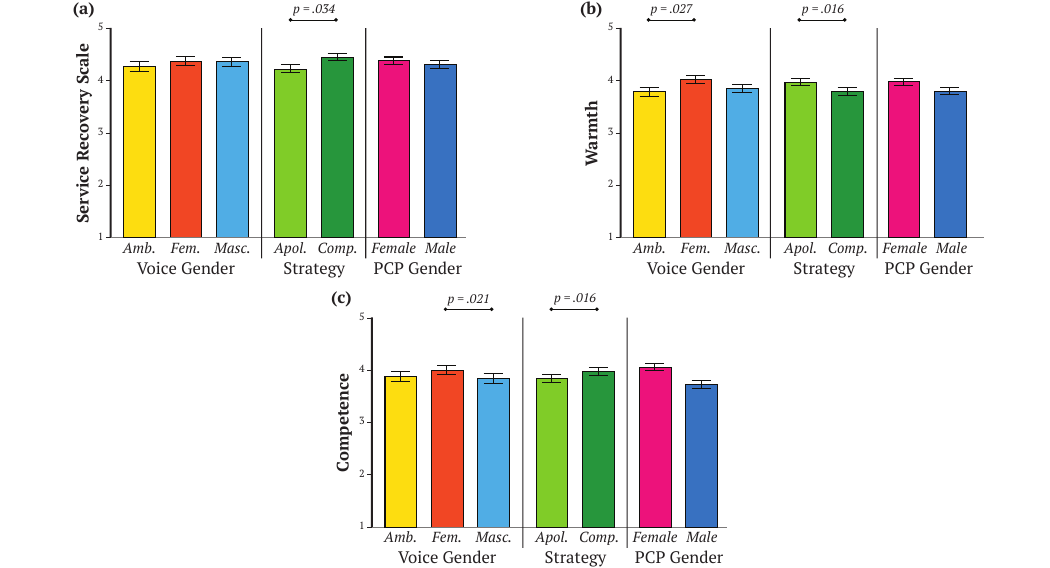}
    \caption{Results of participants' service recovery satisfaction, and perceived warmth and competence of the assistant. Mixed-model repeated measures ANOVAs were conducted to discover effects of voice gender---ambiguous (Amb.), feminine (Fem.), masculine (Masc.)----, mitigation strategy---apology (Apol.), compensation (Comp.)---, and participant (PCP) gender---female, male--- on subjective measures. All pairwise comparisons were conducted using Fisher's LSD method with Bonferroni correction. Error bars represent standard error (SE) and only the significant comparisons ($p<.05$) are highlighted.}
    \Description[Results of participants' service recovery satisfaction, and perceived warmth and competence of the assistant.]{The figure shows results of participants' service recovery satisfaction, and perceived warmth and competence of the assistant. The figure further consists of 3 sub graphs. All are bar plots and y-axis is ranged from 1 to 5. 
    The first of the 3 sub graphs from left to right further consists of 3 bar plots and shows results on service recovery satisfaction which is the y-axis for all three bar plots. 
    First of the 3 bar plots has voice gender on x-axis and shows no significant main effect of voice gender on service recovery satisfaction. the mean for ambiguous is 4.2, and for feminine and masculine sounding assistants is closer to 4.3.
    Second of the 3 bar plots has mitigation strategy on x-axis and shows a significant main effect of mitigation strategy on service recover satisfaction with a p-value of .034. Assistant employing compensation strategy with a mean of about 4.5 is rated better than the assistant that apologizes with a mean of 4.2. 
    Third of the 3 bar plots has participant gender on x-axis and shows no significant  main effect of  participant gender on service recovery satisfaction. Male participants have a mean of 4.2 and female participants have a mean of 4.3.
    The second of the 3 sub graphs from left to right further consists of 3 bar plots and shows results on warmth which is the y-axis for all three bar plots. 
    First of the 3 bar plots has voice gender on x-axis and shows a significant difference between ambiguous sounding and Feminine sounding assistants with a p-value of 0.027. Feminine sounding assistant with a mean of about 4.0 is perceived to be warmer than ambiguous sounding one with a mean of 3.8. Masculine assistant has a mean warmth of 3.9. 
    Second of the 3 bar plots has mitigation strategy on x-axis and shows a significant main effect of mitigation strategy on warmth with a p-value of .016. Assistant employing apology strategy with a mean of about 4.0 is rated better than the assistant that compensates with a mean of 3.8.
    Third of the 3 bar plots has participant gender on x-axis and shows no significant  main effect of  participant gender on assistants' perceived warmth. Male participants have a mean of about 3.8 and female participants have a mean of about 4.0.
    The third of the 3 sub graphs from left to right further consists of 3 bar plots and shows results on competence which is the y-axis for all three bar plots. 
    First of the 3 bar plots has voice gender on x-axis and shows a significant difference between feminine sounding and masculine sounding assistants with a p-value of 0.021. Feminine sounding assistant with a mean of about 4.0 is perceived to be more competent than masculine sounding one with a mean of 3.8. Ambiguous assistant has a mean competence of 3.9. 
    Second of the 3 bar plots has mitigation strategy on x-axis and shows no significant main effect of mitigation strategy on warmth. Assistant employing compensation strategy has a mean of about 4.0 and the assistant that apologizes with a mean of about 3.8.
    Third of the 3 bar plots has participant gender on x-axis and shows no significant  main effect of  participant gender on assistants' perceived competence. Male participants have a mean of about 3.7 and female participants have a mean of about 4.2.
    }
    
    \label{fig:service}
\end{figure*}

\hl{We observed a significant main effect of mitigation strategy of participants' satisfaction of service recovery; in particular, compensation  resulted in better service recovery satisfaction than apology } (Fig. \ref{fig:service} a and Table \ref{tab:subjective-results}).

\subsubsection{Warmth}
\hl{We observed a a significant main effect of voice gender on participants' perception of warmth of agent. The agent with a feminine voice  was perceived warmer than the agent with an ambiguous voice } (Fig. \ref{fig:service} b and Table \ref{tab:subjective-results}).

\subsubsection{Competence}
\hl{We observed  a significant main effect of voice gender  on participants' perception of competence of voice assistant} (Fig. \ref{fig:service} c and Table \ref{tab:subjective-results}). 
\hl{The assistant with a feminine voice was perceived more competent than the assistant with a masculine voice. 
Moreover, there was a significant main effect of the mitigation strategy on perceived competence;  assistants offered compensation were perceived to be warmer than assistants that apologized.}

\begin{table}[htbp]
\centering
\caption{Results of behavioral analysis: Related to error and error mitigation strategy.  $^{\ast}p \leq 0.05$, $^{\ast\ast}p \leq 0.01$, and  $^{\ast\ast\ast}p\leq 0.001$}
\label{tab:behavioral-1}
\begin{tabular}{lll}

\textbf{Participant interruption during error}\\
\hline
Variables & $\chi^2$ & \textbf{$p$}  \\
\hline
\hline
Voice Gender    & $\chi^2(2) = 1.66$ & $.435$  \\
\hline
Participant Gender     & $\chi^2(1) = 4.93$ & $.026^{\ast}$ \\
\hspace{3mm} Female($9.6\%$) < Male($19.8\%$)  \hspace{2mm} $95\%$ CI: $[1.10,4.89]$ \\
\hline
\\
\textbf{Non-verbal reaction to error}\\
\hline
Variables & $\chi^2$ & \textbf{$p$}  \\
\hline
\hline
Voice Gender    & $\chi^2(2) = 3.08$ & $.857$  \\
\hline
Participant Gender     & $\chi^2(1) = 0.03$ & $.857$ \\
\hline
\\
\textbf{Verbal response to error mitigation strategy}\\
\hline
Variables & $\chi^2$ & \textbf{$p$}  \\
\hline
\hline
Voice Gender    & $\chi^2(2) = 0.44$ & $.802$  \\
\hline
Mitigation Strategy     & $\chi^2(1) = 0.09$ & $.766$ \\
\hline
Participant Gender     & $\chi^2(1) = 0.06$ & $.812$ \\
\hline
\\
\textbf{Non-verbal response to error mitigation strategy}\\
\hline
Variables & $\chi^2$ & \textbf{$p$}  \\
\hline
\hline
Voice Gender    & $\chi^2(2) = 1.50$ & $.471$  \\
\hline
Mitigation Strategy     & $\chi^2(1) = 0.01$ & $.937$ \\
\hline
Participant Gender     & $\chi^2(1) = 3.32$ & $.069$ \\
\hline
\end{tabular}

\end{table}

\begin{figure*}[tb]
    \includegraphics[width=\textwidth]{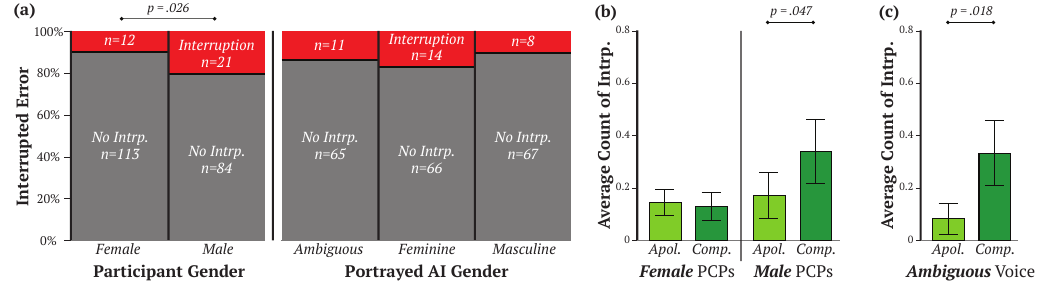}
    \caption{Results of interruption (Intrp.) related behavioral metrics. Generalized Estimating Equations (GEE) was used to fit a logistic regression to discover effects of voice gender, and participant gender on participants' interruption to error. Moreover, a mixed model ANOVA was conducted to study effects of voice gender---ambiguous, feminine, masculine----, mitigation strategy---apology (Apol.), compensation (Comp.)---, and participant (PCP) gender---female, male--- on number of interruptions. All pairwise comparisons were conducted using Tukey's HSD method. Error bars represent standard error (SE) and only the significant comparisons ($p<.05$) are highlighted.}
    \Description[Results of interruption related behavioral metrics]{The figures has results on interruption related user behavior metrics. There are 3 sub graphs. First sub graph from left to right further consists of two mosaic plots. Generalized Estimating Equations (GEE) to fit a repeated measures logistic regression was run for this analysis of interrupted error. First of the two mosaic plots has participant gender on x-axis and interrupted error (dichotomous variable) which shows a significant difference with p-value of .026 between female and male participants. Female participants interrupted 12 times while did not interrupt at all 113 times where as male participants interrupted 21 times and did not interrupt 84 times. The second mosaic plot has portrayed AI gender on x-axis while y-axis is the same which is interrupted error. It shows no significant effect of AI gender on whether participants interrupted during mitigation statement delivery. For ambiguous assistant there was an interruption in 11 of total 76 trials, for feminine assistant in 14 of 80 trials and for masculine assistant in 8 of 75 trials.
    The second and third graphs from left to right show results from a mixed model ANOVA to study effects of our factors on number of interruptions. The y-axis for both plots is average count of interruptions with a range of 0 to 0.8.  The second from left to right has two bar plots. First of the two bar plots is for female participants only and the x-axis is mitigation strategies showing no significant interaction effect between the two and mean for both apology and compensation is around 0.1. For the second of 2 bar graphs, mitigation strategies only for male participants are on x-axis. It shows that apologetic assistants had lesser interruptions with mean about 0.2 than compensating assistants with mean about 0.4 when looking at male participants only. The p-value for this interaction effect is .047. 
    The third and the last graph from left to right has mitigation strategies on x-axis only for ambiguous sounding assistants. Apologetic assistants with mean of about 0.1 were interrupted less than compensating ones with mean of about 0.3 when looking at ambiguous sounding assistants only. The p-value for this interaction effect is .018.}
    \label{fig:interrupted_error}
\end{figure*}

\begin{table}[htbp]
\centering
\caption{Results of behavioral analysis: Relevant to overall interaction.  $^{\ast}p \leq 0.05$, $^{\ast\ast}p \leq 0.01$, and  $^{\ast\ast\ast}p\leq 0.001$}
\label{tab:behavioral-2}
\begin{tabular}{lll}
\textbf{Number of interruptions}\\
\hline
Variables & $F$ & $p$  \\
\hline
\hline
Voice Gender    & $F(2,75.89) = 0.16$ & $.849$ \\
\hline
Mitigation Strategy    & $F(1,38.06) = 3.45$ & $.071$  \\
\multicolumn{2}{l}{\hspace{3mm} Compensation ($M= 3.80, SD=0.77$) < Apology  ($M= 3.98, SD=0.74$)} &\\
\hline
Participant Gender     & $F(1,38.30) = 0.49$ & $.489$  \\
\hline
Participant Gender $\times$ Voice Gender  & $F(2,75.89) = 0.78$ & $.464$  \\
\hline
Participant Gender $\times$ Mitigation Strategy  &$F(1,38.06) = 4.67$ & $.037^{\ast}$ \\
\hspace{3mm} Male, Apology ($M=0.17, SD=0.64 $)  < \\
\hspace{3mm} Male, compensation ($M= 0.34, SD=0.90 $) & & $.047^{\ast}$ \\
\hline
Voice Gender $\times$ Mitigation Strategy  & $F(2,75.29) = 3.80$ & $.027^{\ast}$  \\
\hspace{3mm} Ambiguous, Apology ($M=0.08, SD=0.36$)  \\
\hspace{3mm} < Ambiguous, Compensation ($M=0.33, SD=0.77$) & & $.018^{\ast}$\\
\hline
Participant Gender $\times$ Voice Gender \& Mitigation Strategy  & $F(2,75.29) = 0.58$ & $.562$ \\
\hline
\\
\textbf{Number of other verbal reactions}\\
\hline
Variables & $F$ & \textbf{$p$}  \\
\hline
\hline
Voice Gender    & $F(2,63.99 = 0.04$ & $.960$ \\
\hline
Mitigation Strategy    & $F(1,38.35) = 2.59$ & $.116$  \\
\hline
Participant Gender     & $F(1,37.34) = 0.73$ & $.397$  \\
\hline
Participant Gender $\times$ Voice Gender  & $F(2,63.99) = 0.95$ & $.391$  \\
\hline
Participant Gender $\times$ Mitigation Strategy  &$F(1,38.35) = 0.27$ & $.606$ \\
\hline
Voice Gender $\times$ Mitigation Strategy  & $F(2,66.31) = 2.10$ & $.130$  \\
\hline
Participant Gender $\times$ Voice Gender \& Mitigation Strategy  & $F(2,66.31) = 0.34$ & $.710$ \\
\hline
\\

\textbf{Number of other non-verbal reactions}\\
\hline
Variables & $F$ & \textbf{$p$}  \\
\hline
\hline
Voice Gender    & $F(2,73.65) = 2.80$ & $.067$ \\
\hline
Mitigation Strategy    & $F(1,35.1) = 7.14$ & $.011^{\ast}$  \\
\hspace{3mm}\hspace{3mm} Compensation ($M = 0.40, SD = 0.77$) < Apology  ($M = 0.74, SD = 1.27$)&  &\\
\hline
Participant Gender     & $F(1,38.30) = 1.88$ & $.178$  \\
\hline
Participant Gender $\times$ Voice Gender  & $F(2,73.65) = 4.54$ & $.014^{\ast}$  \\
\hspace{3mm} Male, Masculine ($M = 0.53, SD = 0.90$) < Male, Feminine ($M = 1.08, SD = 1.62$) & & $.029^{\ast}$\\
\hline
Participant Gender $\times$ Mitigation Strategy  &$F(1,35.10) = .69$ & $.411$ \\
\hline
Voice Gender $\times$ Mitigation Strategy  & $F(2,76.79) = 0.19$ & $.828$  \\
\hline
Participant Gender $\times$ Voice Gender \& Mitigation Strategy  & $F(2,76.79) = 4.83$ & $.011^{\ast}$ \\
\hspace{3mm} Male, Masculine, Apology ($M = 0.61, SD = 0.98$) \\
\hspace{3mm} < Male, Feminine, Apology ($M = 1.47, SD = 1.98$) & & $.030^{\ast}$\\

\hline
\end{tabular}

\end{table}

\subsection{Behavioral Analysis}

We did not include the nine trials in which participants did not recognize the error (one trial from a female participant and eight trials from male participants).
As a result, the following analyses included 231 trials.
\hl{Tables} \ref{tab:behavioral-1} and \ref{tab:behavioral-2} \hl{summarize the results of behavioral analysis.}

\subsubsection{Participant interruption during error (Dichotomous variable: 0/Yes, 1/No)}
For this analysis, we did not consider error mitigation as a factor because the participants did not experience the mitigation strategy until after the error was fixed. In other words, whether the participants interrupted the agent during the error was not affected by the mitigation strategy. 
We set up Generalized Estimating Equations (GEE) to fit a 
logistic regression to examine whether gender of voice and participant gender had an effect on whether the participant interrupted the error or not. 
\hl{The model estimated that the odds for male participants ($19.8\%$) interrupting the assistant after realizing the error has occurred while assistant was still talking are approximately 2.32 times higher than female participants ($9.6\%$) interrupting the assistant} (Fig. \ref{fig:interrupted_error} a and Table \ref{tab:behavioral-1}).

\begin{figure*}[t]
    \includegraphics[width=\textwidth]{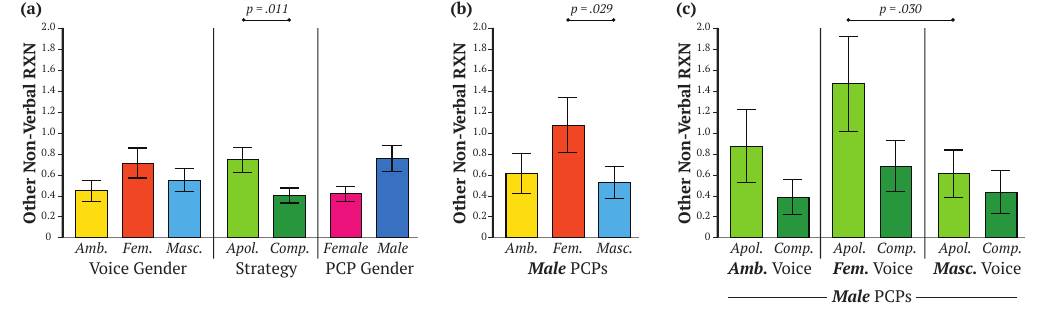}
    \caption{Results of number of other non-verbal reactions (RXN). Mixed-model repeated measures ANOVA was conducted to discover effects of voice gender---ambiguous (Amb.), feminine (Fem.), masculine (Masc.)---, mitigation strategy---apology (Apol.), compensation (Comp.)---, and participant (PCP) gender---female, male--- on number of other non-verbal reactions. All pairwise comparisons were conducted using Tukey's HSD method. Error bars represent standard error (SE) and only the significant comparisons ($p<.05$) are highlighted.}
    \Description[Results of number of other non-verbal reactions]{There are four bar charts for mixed model repeated measures ANOVAs to show effects of our factors on number of other non-verbal reactions. The range of y-axis is 0 to 1.6 for all graphs. First bar chart from left to right further consists of 3 bar charts. The first among the three bar charts shows voice gender on the x-axis to present that there is no significant main effect of gender of AI voice on number of other non-verbal reactions. The mean for ambiguous sounding assistant is around 0.4, female sounding assistant is around 0.6 and male one is around 0.5. 
    The second of the 3 bar charts shows mitigation strategy on x-axis to present that there is significant main effect of gender on the number of non-verbal reactions, p-value is .025 where mean for apology is about 0.6 and is higher than mean for compensation which is about 0.4. the third of the 3 bar charts shows participant gender on the x-axis and shows no significant effect of participant gender on count of other non-verbal reactions. The mean for female participants is around 0.4 while mean for male participants is around 0.6. The second bar chart from left to right further consists of one bar chart for showing interaction of voice gender and participant gender on the number of other non-verbal reactions. It only shows plot for error mitigation strategies for male participants. Thus, the x-axis is ambiguous with mean of about 0.6, female with mean of about 0.9 and male assistants with mean of about 0.4 for male participants only. There is only a significant difference between female and male assistants with female having higher mean than male assistants while looking at male participants only. The p-value for this interaction is .032. 
    The third bar chart from left to right further consists of 3 bar charts. It only shows all the graphs for male participants only. The first of the three bar charts shows mitigation strategies for male voice assistants only, on x-axis. There is no significant difference between apology with a mean of about 0.8 and compensation strategy with a mean of about 0.6 when it comes to number of non-verbal reactions. The second of the three bar charts shows mitigation strategies for only female voice assistants on x-axis while the third of the three bar charts only shows mitigation strategies for male voice assistants. There is only one significant interaction effect between these two bar charts which is that male participants react non-verbally more to apologetic female assistants with mean of about 1.2 than apologetic male assistants with mean of about 0.4. The p-value for this interaction is .042. Mean for female voice assistants that compensate is about 0.6 and for male voice assistant that compensated is about 0.4.}
    \label{fig:count-non-verbal}
\end{figure*}

\subsubsection{Non-verbal reaction to error (Dichotomous variable: 0/Yes, 1/No)}
For this measure, we also did not considering error mitigation as factor as well because the participants did not interact with this manipulation till after error was fixed. Hence, whether the participants reacted non-verbally to the agent during the error was not affected by the mitigation strategy. 
We set up GEE to fit a logistic regression to examine whether voice gender and participant gender influenced non-verbal reaction to error. \hl{We saw no significant results.}

\subsubsection{Verbal response to error mitigation strategy (Dichotomous variable: 0/Yes, 1/No)}

We set up GEE to fit a 
logistic regression to examine whether voice gender, mitigation strategy, and participant gender influenced verbal response to error mitigation strategy. \hl{We did not see any significant effects.}

\subsubsection{Non-verbal response to error mitigation strategy (Dichotomous variable: 0/Yes, 1/No)} 
We set up GEE to fit a 
logistic regression to examine whether voice gender, mitigation strategy, and participant gender influenced non-verbal response to error mitigation strategy. \hl{No significant effects were observed.}

\subsubsection{Number of interruptions}
\hl{A mixed repeated measures ANOVA with voice gender and mitigation as within-subjects factors, and participant gender as a between-subjects factor revealed a significant interaction effect of participant gender and mitigation on count of interruptions. Post-hoc comparisons using Tukey's HSD revealed that there were more interruptions between male participants and agents compensating than the agents apologizing, regardless of assistant's gender} (see Fig. \ref{fig:interrupted_error} b and Table \ref{tab:behavioral-2}). 
\hl{Moreover, a significant interaction effect of voice gender and mitigation on count of interruptions was also observed. For ambiguous sounding agents, there were more interruptions when they compensated  rather than apologized for the errors} (see Fig. \ref{fig:interrupted_error} c and Table \ref{tab:behavioral-2}).

\subsubsection{Number of other verbal reactions}
Using mixed model repeated measures ANOVA with voice gender and mitigation strategy as within-subjects factors, and participant gender (male, female) as a between-subjects factor, \hl{we observed no significant effects.}

\subsubsection{Number of other non-verbal reactions}
Fig. \ref{fig:count-non-verbal} and Table \ref{tab:behavioral-2} summarize our results for number of other non-verbal reactions.
\hl{Using mixed model repeated measures ANOVA with voice gender and mitigation strategy as within-subjects factors, and participant gender (male, female) as a between-subjects factor, we observed a significant main effect of mitigation strategy on the number of non-verbal responses. Participants reacted non-verbally more to agents in the apology condition than in the compensation condition.
There was a significant interaction effect between participant gender and the voice gender of AI. Tukey's HSD shows that male participants reacted non-verbally more to agents with feminine  voice than to agents with masculine voices.
Similarly, a significant interaction effect was found between participant gender, mitigation strategy, and the voice gender of AI. Tukey's HSD shows that male participants reacted non-verbally more to apology coming from the agent with feminine voice than from the agent with masculine voice.}

\section{Discussion}\label{sec:discussion}
Mitigating AI errors during recovery from service failures is essential to restoring and maintaining human-AI collaboration. AI agents in tasks of assistance nature are mostly defaulted to manifesting feminine voices and characteristics, which may reinforce biases because of association of gender stereotypes. Users' gender may also influence perceptions of gendered AI agents. Gender ambiguous voices have been explored as a way to mitigate these biases. This work explored how \textit{error mitigation strategy} (apology vs. compensation), \textit{gender of AI assistant voice} (feminine vs. ambiguous vs. masculine), and participant's gender (female vs. male) may affect participants' satisfaction with the service recovery, perception of, and their behavior around voice assistants \hl{during an assistive task exemplified as an online shopping task. }
Below, we discuss our results, their implications, limitations of this study, and future work.

\subsection{Apology Communicates Warmth and Compensation Leads to User Satisfaction}
\label{sec:discussiion_warmth}
Our hypothesis 1 predicts that apology during recovery from errors by voice assistants portrays the assistant as warmer while offered compensation creates greater service recovery satisfaction and perceptions of competence. Our results support this hypothesis
(Fig. \ref{fig:service} and Table \ref{tab:subjective-results}). 
\hl{Prior work has demonstrated that humans exhibit social reactions in response to conversational agents when these agents exhibit social personality such as engaging in small talk, displaying gender, and age characteristics, using gestures, or expressing facial expressions} \cite{feine2019taxonomy}. \hl{In our study, we see that} voice assistants that were apologetic 
---exhibiting a social personality--- elicited more non-verbal reactions than the ones that offered compensation ---exhibiting a utilitarian personality--- (Fig. \ref{fig:count-non-verbal} a and Table \ref{tab:behavioral-2}). \hl{For example, one participant laughed 13 times when interacting with assistants that apologized while only 5 times for assistants that compensated. }
These non-verbal reactions such as smile, laugh, and nod are reflective of social reactions.
\hl{Since, apologizing or offering compensation can be seen as indications of distinct personalities or traits communicated through speech, }
\hl{ \textbf{the level of non-verbal user reactions can serve as an indicator of users' perceptions of voice assistant's personality}.}

\hl{However, further looking into the reactions, their valence may vary for different participants. For instance, one participant reacted a total of six times to apologetic agents in which two of the reactions were ``nodding'', two were ``shaking head'', and two were ``hands in the air'' gestures indicating confusion. While both of his reactions to compensating agents were  ``hands in the air'' gestures indicating confusion. The apologetic agents were able to elicit a greater variety of expressions with both positive and negative connotations. 
Though may be participant dependent, social reactions can be informative of users' perceptions of agent's personality traits such as warmth and competence. 
This observation has design implications for using social reactions to understand users' perceptions of assistants and is similar to the prior findings illustrating that social reactions from people are indicative of their surprise and experiences of unexpected events (e.g., errors) during their interactions with robots} \cite{stiber2022modeling,stiber2023using}.
\hl{However, further research is needed to study how the valence of non-verbal reactions may be used to model and estimate users' perceptions of AI agents, which could then be used to adapt the agents' behaviors towards achieving desirable interaction outcomes. }

\subsection{Portrayed Voice Gender Influences User Perceptions of the Voice Assistant}
\label{sec:female-job}
Our hypothesis 2 predicts that masculine voice assistants are viewed as more competent while feminine voice assistants are considered warmer. Our results do not support this hypothesis since we observe that the feminine assistants are perceived as more competent than masculine assistants (Fig. \ref{fig:service} and Table \ref{tab:subjective-results}). We speculate that \hl{this preference of} feminine voice assistants \hl{in the context of online shopping assistance may be influenced by the broader societal stereotypes associating women with competence in serving supportive roles in assistance-type tasks}
\cite{powell2002gender, koenig2011leader}. 
\hl{It is noteworthy that commonly available voice assistants such as Siri, Alexa, and Cortana are marketed as ``personal assistants'' and typically employ feminine voices }
\cite{un2019report}; \hl{such de facto voice setting has shown to attract abusive behavior} \cite{curry2019crowd, curry2020conversational}. 
\hl{Such stereotypes have been translated to the preference of feminine voice assistants in tasks that are ``perceived'' as domestic chores such as shopping for groceries, schools supplies, and more}
\cite{lee2019voice}. 
\hl{While digital personal assistance should not be inherently associated with a specific gender, when a gendered voice (e.g., feminine voice) is assigned to a voice assistant, user perceptions of the online shopping task may align with the gender stereotypes and biases seen in prior work}  \cite{lee2019voice,ahn2022effect,damen2019designing,tay2014stereotypes}.
\hl{Hence, the association of competence with feminine agents for assistance-type tasks found in our study possibly hint at implicit preferences and biases driven by gender stereotypes.}




The text-to-speech systems are evolving to generate increasingly realistic and humanlike voices for AI agents \cite{seaborn2021voice}.
\hl{In this context, our study revealed that } 
 the feminine voice is perceived to be more humanlike ($M=3.36$, 1 being robotic and 5 being humanlike) as compared to masculine ($M= 2.96$) and ambiguous one ($M=2.59$). 
We speculate 
that the higher perception of humanlikeness for feminine voices may again be attributed to users' familiarity with feminine assistants in real life, such as Alexa, Siri, and Bixby.
\mytc{However, as voice assistants become more humanlike due to addition of social cues to their behavior} \cite{bucher2022form}, biases encountered by these assistants \hl{could potentially} reinforce gender preferences and biases in actual human interactions. 
Therefore, \textbf{designers and developers must carefully consider the portrayed gender of voices while designing AI agents for specific tasks to support people effectively while minimizing potential negative consequences in both human-agent and human-human interactions}.

\subsection{Men Show More Biased Behavior}
Our results highlight the impact of user gender on their behavior towards gendered voice assistants while repairing human-agent relationship using different mitigation strategies. Male participants generally reacted non-verbally (socially) more to feminine assistants than masculine ones (Fig. \ref{fig:count-non-verbal} b and Table \ref{tab:behavioral-2}). \hl{For example, a male participant smiled at feminine assistants 
more than at masculine assistants. Similarly, another male participant presented various types of reactions (two nodes, one smile, and one frown) to feminine assistants whereas he only frowned once at the masculine assistants.} 

Moreover, male participants registered more non-verbal reactions to ``apologetic'' agents when they sounded feminine as opposed to masculine (Fig. \ref{fig:count-non-verbal} c and Table \ref{tab:behavioral-2}). \hl{For instance, a male participant laughed six times when interacting with feminine sounding apologetic assistant while only one time to the masculine counterpart. Similarly, another male participant nodded twice, smiled once, and raised an eyebrow (frowned) once in reaction to the feminine sounding apologetic assistant, whereas no reaction was given to the masculine sounding apologetic assistant.} \hl{As previously discussed in Section} \ref{sec:discussiion_warmth}\hl{, we see that the valence and variety of reactions may vary depending on the individual participant. Future research could explore how such diverse reactions contribute to the reflection of biased behavior in assistance scenarios.}

Prior work has also shown similar discrepancy in behavior between men and women. 
\hl{Men found robots with feminine voices more trustworthy and engaging, leading them to donate more money. On the other hand, women show no preference between robots with feminine or masculine voices }
\cite{siegel2009persuasive}. 
\hl{Consistent with these prior findings, our results imply that male participants engage more with feminine assistants, especially during apologetic interactions which \textbf{highlights the importance of considering user gender and its potential impact on interactions with gendered voice assistants, especially considering that these assistants will inevitably make mistakes in their interactions with people in complex, real-world settings. }}


\hl{We further see biased behavior towards voice assistants by men---}
regardless of the gender of voice assistants and error mitigation strategy, male participants were found to interrupt the assistants on realization of error before the assistant stopped talking almost twice as many times as female participants would  (Fig. \ref{fig:interrupted_error} a and Table \ref{tab:behavioral-1}). 
\hl{These interruptions can be seen as impolite behavior}
\cite{hutchby2008participants},
\hl{possibly stemming from implicit power dynamics where men are typically in authoritative roles }
\cite{powell2002gender}. 
\hl{Previous research in human-agent interactions, has also indicated that men show less inclination to assist the agent, such as in the case of an AI study companion }
\cite{lubold2016effects}.
\hl{Such biased behavior towards voice assistants by men highlights the need for further examination of implicit power dynamics and their impact on human-agent interactions.}

We also observe that that male participants ($M=3.17$), as opposed to their female counterparts ($M=3.45)$, overall, perceived the voice assistants to be more feminine regardless of the portrayed gender of the voice and mitigation strategy (Fig. \ref{fig:gender} a and Table \ref{tab:characteristics-perceptions-results}). 
Although this perception may arise from a greater familiarity with ``feminine commercial voice assistants'' as discussed in section \ref{sec:female-job}, it can still have a negative impact on interactions with AI assistants.
Hence, it is of utmost importance to further understand impacts of association of any gender to voice assistants, and interaction of agents' and users' gender before continuing to create gendered voice assistants aimed to assist and support people. 

Despite male participants interrupting the voice assistants more during error, we see that the interruptions between the male participants and the voice assistants were fewer in number when the assistants were apologizing for errors instead of compensating the participants (Fig. \ref{fig:interrupted_error} b and Table \ref{tab:behavioral-2}). A possible explanation for this finding is that it may be considered rude to interrupt while someone is apologizing. Thus, voice assistants can potentially utilize apologies to deter people from interrupting, thereby fostering more polite, smooth, and respectful interactions.  This finding implies that \textbf{we can tailor agent behavior such as using apology as a mitigation strategy to elicit positive attitude from people in face of imperfect interactions}.

\subsection{Gender Ambiguous Voices as a Plausible Means to Reduce Biases}

\hl{Gender ambiguous voices (e.g., ``Q, the genderless''\footnote{\url{ https://www.genderlessvoice.com/}}) is presented as a tool to culminate gender stereotypes and biases in prior work on voice assistants} \cite{tolmeijer2021female}. 
Therefore, 
in addition to our hypotheses 2 and 3, we explore if an ambiguous voice can eliminate or reduce the effects of stereotypes associated with gendered voices for AI agents. Our results show that ambiguous sounding assistants are perceived less warm compared to feminine ones (Fig. \ref{fig:service} b and Table \ref{tab:subjective-results}), suggesting that ambiguous voices can potentially be employed to isolate gender stereotypes around warm personality (feminine agents or robots are considered warmer than masculine ones \cite{ahn2022effect,nomura2017robots}) for voice assistants. This finding contradicts a previous study where feminine and ambiguous voices are not disassociated with stereotypical feminine traits such as family-orientation, delicacy, and sensitivity \cite{tolmeijer2021female}. 
\hl{The distinct perception of warmth of our gender-ambiguous voice, as compared to feminine one, indicates the potential for using ambiguity of gender in voice to create less biased voice interactions with AI agents.}

Looking into behavioral measures, the interruptions between users and ambiguous voice assistants were more frequent when the assistants compensated instead of apologizing for the error (Fig. \ref{fig:interrupted_error} c and Table \ref{tab:behavioral-2}).
Fewer interruptions generally reflect increased politeness \cite{hutchby2008participants}. 
\hl{Therefore, by employing a relational strategy like apologizing, ambiguous assistants were able to elicit politeness and courtesy from users without significant interruptions, despite being perceived as more robotic than gendered assistants. }
Therefore, a design implication of our work is that \textbf{ambiguous voices for AI agents have the potential to maintain certain benefits, such as promoting politeness and attentive listening, while reducing the association with stereotypical feminine traits (such as warmth and humanlikness) when the agent's role is assistive in nature.}

\hl{Our study additionally shows that ambiguous sounding assistants are perceived to be the most robotic when compared to feminine and masculine ones. This result, in combination with prior work}
showing that majority of people preferred a robotic voice ($32.4\%$), followed by feminine ($20.6\%$) and masculine ($11.8\%$) voice for conversational assistants \cite{curry2020conversational}, \hl{highlights potential of gender ambiguous voices being welcomed by users. However, \textbf{further research is needed to establish the effectiveness of ambiguous voices in a variety of assistance-type tasks; this work only serves as an initial exploration in this direction}.}


\subsection{Limitations and Future work}
This work has a few limitations that can inform future work. 
First, online shopping with voice assistants is novel to most people because of limited capabilities and hindrance in adoption of the smart speakers for functional tasks with higher risks than putting alarms and reminders, controlling smart appliances, information retrieval, and playing music. 
Second, the experiment was conducted in a lab setting where participants might not be fully comfortable to interact freely with the speaker. 
\hl{Future work should investigate the integration of voice assistants into people's personal physical spaces to better understand the organic interaction between humans and agents in commonplace and everyday tasks.}

Third, limited voices were explored for masculine, feminine and ambiguous agents in this study. Future work can focus on understanding various characteristics of voice generation towards creation and testing of genderless and ambiguous voices. Moreover, there are other aspects, such as personality, names assigned, and representations in form of color and appearance
, to design of voice assistants than just voice that may influence perception of their gender \cite{sutton2020gender}. 

Fourth, our participants do not represent the full spectrum of gender, age, ethnicity, and education. Thus, our findings and implications should be interpreted with this context in mind. 
Specifically, due to large imbalance in gender of participants, we could not include non-binary participant gender in our analyses. Therefore, effects of gendered and ambiguous voices on non-binary population is unexplored. Future research should aim to recruit participants representative of all genders to understand interaction of users' and assistants' gender. 
Moreover, age of users may affect how they interact with voice assistants as these assistants are likely to make more errors when interacting with older adults due to the diminished ability in their speech such as difficulty in constructing a structured sentence for a command \cite{kim2021exploring}. 

Overall, future research should explore voice-based interactions with virtual assistants to assist people in realistic settings and how various aspects of agent design such as voice, personality, and style of speech may be manipulated to reduce biases caused due to the association of stereotypes based on humanlike characteristics, such as race and gender, projected onto AI agents.

\section{Conclusion}
\hl{The growing use of intelligent voice assistants, their susceptibility to errors, and the potential reinforcement of gender biases in user interactions motivated us to study how gender portrayal through gendered voice influences error mitigation strategies employed by these voice assistants. 
Our results reveal that people exhibit higher social engagement, as evidenced by non-verbal reactions, when interacting with voice assistants that apologize rather than compensate for errors, possibly because apologetic voice assistants are perceived as warmer while compensating ones are viewed as more competent.
Moreover, we observe that men prefer interacting with feminine voice assistants that apologize for mistakes during assistive tasks. Furthermore, our study reveals a notable gender difference in interruption patterns---men interrupted voice assistants more frequently than women when errors occurred, indicating potential sociocultural and power dynamics at play.
Apart from illustrating biases in users' perceptions and behaviors towards gendered voice assistants, our study highlights the potential of using ambiguous voices to mitigate the association of stereotypical feminine traits, such as warmth, with AI agents designed to assist people. 
This work takes a step towards designing characteristics and behavior of voice assistants to foster interactions that are engaging, while simultaneously addressing potential gender biases, with the ultimate goal of enhancing the overall user experience.}

\bibliographystyle{ACM-Reference-Format}
\bibliography{references}
\end{document}